\documentclass[referee]{raa}            

\usepackage{graphicx,times}             
	\usepackage{amsmath}
    \usepackage{color}

\begin{document}

   \title{Evolution of the Coronal Magnetic Configurations Including a Current-Carrying Flux Rope in Response to the Change in the Background Field
}

   \volnopage{Vol.0 (200x) No.0, 000--000}      
   \setcounter{page}{1}          

   \author{Hong-Juan Wang,
      \inst{1}
   \and Si-Qing Liu
      \inst{1}
   \and Jian-Cun Gong
      \inst{1}
   \and Jun Lin
      \inst{2}
   }

   \institute{Center for Space Science and Applied Research, Chinese Academy of Sciences,
Beijing 100190, China \\
        \and
             Yunnan Observatory, Chinese Academy of
              Sciences, Kunming, Yunnan 650011, China
   }

   \date{Received~~; accepted~~}

\abstract{ 
We investigate equilibrium height of the flux rope, and its internal equilibrium in a
realistic plasma environment by carrying out numerical simulations of the evolution
of systems including a current-carrying flux rope. We find that the equilibrium
height of the flux rope is approximately a power-law function of the relative strength
of the background field. Our simulations indicate that the flux rope can escape more
easily from a weaker background field. This further confirms the catastrophe in the
magnetic configuration of interest can be triggered by decrease of strength of the
background field. Our results show that it takes some time to reach internal equilibrium
depending on the initial state of the flux rope. The plasma flow inside the flux rope due to
the adjustment for the internal equilibrium of the flux rope remains small and does not last very
long when the initial state of the flux rope commences from the stable branch of the theoretical
equilibrium curve. This work also confirms the influence of the initial radius of flux rope on its evolution,
the results indicate that the flux rope with larger initial radius erupts more easily.
In addition, by using the realistic plasma environment and much higher resolution in our simulations,
we notice some different characters compared to previous studies in Forbes (1990).
}

   \authorrunning{Wang et al. }            
   \titlerunning{A Catastrophe Model for Coronal Mass Ejections }  

   \maketitle

\section{Introduction}

The most intense energetic activity in the solar system may be
solar coronal mass ejection (CME). During the process, a large
number of magnetized energetic plasmas (with mass of up to
$10^{16}$ g and energy of $10^{32}$ erg) are ejected into the
interplanetary space within a short timescale, and hence disturb
spatial and planetary magnetic field and significantly affect
satellite operation, aviation power, human space exploration,
communication and so on (Chen et al. 2002, Schwenn 2006,
Pulkkinen 2007, Lin 2007, Chen 2011, Mei et al. 2012b,
Shen et al. 2013, Li et al. 2013, and references therein).

It is generally believed that two processes would have
been involved in the intense solar activities and eruptions.
The first one is called storage phase, in which the magnetic
flux transported from the photosphere is slowly accumulated
into the corona, leading to the gradual increase of the
magnetic energy in the corona. The timescale of the magnetic
storage phase is typically several days, so this phase can be
considered as evolving through a series of equilibria in
quasi-static processes. When the stored magnetic energy surpasses
the critical values, the equilibrium will be broken, and the
eruptive phase, i.e. the second process, will occur. The system
in this phase will expand promptly in a dynamic timescale of a
few minutes (i.e. in Alfv$\acute{e}$n time scales) due to loss
of balance.
Transition from the quasi-static evolution to the dynamic phase
constitutes the so-called catastrophe (e.g. Forbes \& Isenberg 1991;
Isenberg et al. 1993;  Forbes \& Priest 1995; Forbes 1994, 2000;
Lin \& Forbes 2000; Lin et al. 2006; Yu 2012, and references therein).

It is well known that the motion of the
photospheric material transports the energy
to the coronal magnetic field for driving
the eruption, so triggering and the consequent
propagation of CMEs are governed by the changes
in the photosphere. Recently, based on MHD
simulation, some authors (Zhang et al. 2011,
Welsch et al. 2009, Kusano et al. 2012,
Kliem et al. 2013, and references therein)
investigated how the physical features in the
photosphere influence the evolution in the coronal
magnetic configuration as well we the initiation
of CMEs. However, without detailed observations
and theoretical simulations, the determinations
for the onset of CMEs still remain unclear.

The decay of the background magnetic field may be a cause
to deviate the CME progenitor structure from the equilibrium,
as shown in Isenberg et al. (1993) and Lin et al. (1998). This
decay could be a consequent of the magnetic diffusion that
leads to the formation, as well as the eruption, of flux rope
(Mackay \& van Ballegooijen 2006). Gradually decreasing the
background field may also cause the state of the flux rope to
transit catastrophically from the old equilibrium to the new
one (Forbes \& Isenberg 1991;Isenberg et al. 1993).
Lin et al. (1998) analytically extended this work to include
the curvature force that creates an additional outward force,
and further realized that, in addition to the impact of the
background field, the radius of flux rope plays an important
role in its eruption. However, these results are constrained by the analytical method.
Although Wang et al. (2009) and Mei et al. (2012a) numerically
investigated the evolution of flux rope, the initial
distribution of the plasma density in the
background field in their simulations are a little bit far
from the realistic case.

We in this paper will numerically investigate the
evolution of the magnetic configuration and the
current-carrying flux rope with the consideration
of the gravitational stratification effect and
more realistic distribution of the plasma density
in the background field, which is
crucial for the generation of CMEs, the understanding
of the catastrophe model for CMEs, and therefor can
allow us to further study the solar-terrestrial
relationship. In addition, a number of numerical experiments have
also been carried out to study how the variation of
the background fields triggers the eruption of the flux
rope and the influence of the radius of the flux rope
on the eruption in detail.
In \S 2, we describe the physical model, formulae and
numerical approaches. The numerical results are presented
in \S 3. We make a discussion and draw conclusions in \S 4.

\section{Physical Model and Numerical Method}

We consider that the prominence or the filament floating in the corona can be represented by
a current-carrying flux rope, and the photospheric background field is represented by a
line dipole below the photosphere. We assume a two-dimensional magnetic configuration in the
semi-infinite $x$-$y$ plane in the Cartesian coordinates. In the coordinates, $y=0$ is assumed
to be the boundary between the photosphere and the chromosphere, $y>0$ corresponds to the
chromosphere and the corona.
The location of the flux rope in our simulations is assumed to be $y=h$ above the boundary $y=0$,
and the depth of the photospheric background field is $y=-d$ below $y=0$.
The empirical atmosphere model described in Sittler \& Guhathakurta (1999) (hereafter S\&G) is used for
the initial background field density $\rho_{0}(y)$.
The evolution of the magnetic system should satisfy the following ideal magnetohydrodynamic (MHD) equations:
\begin{eqnarray}
\begin{split}
&\frac{\textrm{D}\rho}{\textrm{D}t}+\rho\nabla\cdot \textbf{v}=0,
\label{eq:cont}\\
&\rho\frac{\textrm{D}\textbf{v}}{\textrm{D}t}=-\nabla
p+\frac{1}{c}\textbf{J}\times \textbf{B}+\rho\frac{GM_{\bigodot}}{(R_{\bigodot}+y)^{2}}, \label{eq:momen}
\\
&\rho\frac{\textrm{D}}{\textrm{D}t}(e/\rho)=-p\nabla\cdot
\textbf{v},\label{eq:energy}\\
&\frac{\partial \textbf{B}}{\partial t}=\nabla\times(\textbf{v}\times
\textbf{B}),\label{eq:induc}\\
&\textbf{J}=\frac{c}{4\pi}\nabla\times \textbf{B},\label{eq:current}\\
&p=(\gamma-1)e,\label{eq:p1}\\
&p=\rho kT/m_{p}, \label{eq:p2}
\end{split}
\end{eqnarray}
where $\textbf{B}$ represents the magnetic field, $\textbf{J}$ the electric current density,
$\rho$ the mass density, $\textbf{v}$ the velocity of the flow, $p$ the gas pressure, $e$
the internal energy density, $\gamma$ the ratio of specific heats, $G$ the gravitational
constant, $M_{\odot}$ the solar mass, $R_{\odot}$ the solar radius, $m_{p}$ the proton mass.
Equations in (\ref{eq:cont}) are numerically solved by using the ZEUS-2D MHD code described
in Stone \& Norman (1992a, 1992b,1992c).

The magnetic configuration in our simulations is composed of the current-carrying flux rope,
the image of the current inside the flux rope, and the background magnetic field. We assume
that the background field is generated by a line dipole below the bottom of the chromosphere
(Forbes 1990; Wang et al. 2009). The relative strength of the dipole field $M$ can be defined
by a dimensionless parameter $M=m/(Id)$, which is related to the ratio of the strength of the
dipole field $m$ and the product of the filament current $I$ and the depth $d$ of the dipole
field.

The initial magnetic configuration from which the eruption occurs is given by
\begin{eqnarray}
B_{x}&=&B_{\phi}(R_{-})(y-h_{0})/R_{-}-B_{\phi}(R_{+})(y+h_{0})/R_{+}\nonumber\\
&-&B_{\phi}(r+\Delta/2)Md(r+\Delta/2)[x^{2}-(y+d)^{2}]/R^{4}_{d},
\label{eq:Bx}\\
B_{y}&=&-B_{\phi}(R_{-})x/R_{-}+B_{\phi}(R_{+})x/R_{+}\nonumber\\
&-&B_{\phi}(r+\Delta/2)Md(r+\Delta/2)2x(y+d)/R^{4}_{d},\label{eq:By}
\end{eqnarray}
with
\begin{eqnarray*}
R^{2}_{\pm}&=&x^{2}+(y\pm h_{0})^{2},\\
R^{2}_{d}&=&x^{2}+(y+d)^{2}.
\end{eqnarray*}

As for the initial background plasma density $\rho_{0}(y)$, we use an empirical atmosphere S\&G model:
\begin{eqnarray}
\begin{split}
&\rho_{0}(y)=\rho_{00}f(y),\\
&f(y)=a_{1}z^{2}(y)e^{a_{2}z(y)}[1+a_{3}z(y)+a_{4}z^{2}(y)+a_{5}z^{3}(y)],\\
&z(y)=\frac{R_{\odot}}{R_{\odot}+y}, \label{eq:rho0}
\end{split}
\end{eqnarray}
where $\rho_{00}=1.672\times 10^{-13}$ g~cm$^{-3}$,
which is about one order of magnitude smaller
than that in our previous work ($\sim
10^{-12}$~g~cm$^{-3}$, Wang et al. 2009), and $a_{1}=0.001272$,
$a_{2}=4.8039$, $a_{3}=0.29696$, $a_{4}=-7.1743$, $a_{5}=12.321$.
The height $y$ is measured from the surface of the Sun. Equations
(\ref{eq:rho0}) give a slowly decreasing density distribution $f(y)$
for the atmosphere in the lower corona.
This density distribution was supported by the radio observations
of type III bursts over wide frequency band of a few kHz to 13.8 MHz
(Leblanc et al. 1998; Lin 2002).
The density model considered in this work is more realistic than that
used in previous work (Wang et al. 2009; Mei et al. 2012a).

For the initial background atmosphere, there is a balance between pressure
gradient of gas and the gravity
\begin{eqnarray}
\nabla p_{0}(y)=-\rho_{0}(y)\frac{GM_{\odot}}{(R_{\odot}+y)^{2}}. \label{eq:p0}
\end{eqnarray}
From equations (\ref{eq:rho0}) and (\ref{eq:p0}), we can get the relation
between the initial background pressure $p_{0}(y)$, and the temperature
distribution $T_{0}(y)$ as follows
\begin{eqnarray}
p_{0}(y)=\frac{\rho_{0}(y)}{m_{p}}kT_{0}(y), \label{eq:T0}
\end{eqnarray}
where $k$ is the Boltzmann constant.

Subsequently the initial total pressure, including the gas pressure and the
magnetic pressure, and the mass density can be written as
\begin{eqnarray}
\begin{split}
&p=p_{0}-\int^{\infty}_{R_{-}}B_{\phi}(R)j(R)dR,\\
&\rho=\rho_{0}(p/p_{0})^{1/\gamma}. \label{eq:p3}
\end{split}
\end{eqnarray}

$B_{\phi}(R)$ in equations (\ref{eq:Bx}), (\ref{eq:By}) and (\ref{eq:p3}) is
determined by the electric current density distribution $j(R)$
inside the flux rope, and reads as
\begin{eqnarray}
B_{\phi}(R)&=&-\frac{2\pi}{c}j_{0}R, \hspace{2mm} {\mbox for
}\hspace{2mm}0\leq R\leq r-\Delta/2,
\nonumber\\
B_{\phi}(R)&=&-\frac{2\pi j_{0}}{c R}\left\{\frac{1}{2} \left(r -
\frac{\Delta}{2}\right)^{2} - \left(\frac{\Delta}{\pi}\right)^{2} +
\frac{1}{2}R^{2} + \frac{\Delta R}{\pi}
\sin\left[\frac{\pi}{\Delta}\left(R-r+\frac{\Delta}{2}\right)\right]
\right.
\nonumber\\
&+& \left. \left(\frac{\Delta}{\pi}\right)^{2}
\cos\left[\frac{\pi}{\Delta}\left(R-r+\frac{\Delta}{2}\right)\right]\right\},
\hspace{2mm}{\mbox for }\hspace{2mm} r- \Delta/2 < R <r + \Delta/2,
\nonumber\\
B_{\phi}(R)&=&-\frac{2\pi j_{0}}{c
R}\left[r^{2}+(\Delta/2)^{2}-2(\Delta/\pi)^{2}\right],
\hspace{2mm}{\mbox for }\hspace{2mm} r+\Delta/2\leq R<\infty;
\nonumber\\
j(R)&=& j_{0}, \hspace{2mm}{\mbox for }\hspace{2mm} 0\leq R\leq
r-\Delta/2,
\nonumber\\
j(R)&=&\frac{j_{0}}{2}{\cos[\pi(R-r+\Delta/2)/\Delta]+1},\hspace{2mm}
{\mbox for } \hspace{2mm}r-\Delta/2<R<r+\Delta/2,
\nonumber\\
j(R)&=&0, \hspace{2mm}{\mbox for }\hspace{2mm} r+\Delta/2\leq
R<\infty.
\end{eqnarray}

We take the computational domain to be $(-4L, 4L)\times (0, 8L)$ with
$L=10^{5}$ km, and the grid points to be $800\times 800$.
A line-tied condition is applied to the bottom boundary at $y=0$, while
the open boundary condition is used for the other three.
The initial values of the parameters in our simulations are listed
in Table \ref{tbl:1}.

\begin{table}
\caption{The initial values for the important parameters in the numerical experiments.}
\begin{tabular}{llll}
\hline
$\rho_{00}=1.672\times10^{-13}$ g~cm$^{-3}$ & $T_{00}=10^{6}$ K & $j_{00}=1200$ statamp~cm$^{-2}$ & $\gamma=5/3$\\
\hline
\end{tabular}
\label{tbl:1}
\end{table}

\begin{table*}
\center
\caption{Parameters and their values for different cases in the simulations. }
\begin{tabular}{lcccccc}
\hline
Case& $M$   & $d$ (km)  &  $h_{0}/d$     &   $r_{0}/d$ &   $r_{0}/\Delta$   &\\
\hline
1&  2.25    &  $0.125\times10^{5}$     &   0.5   &   0.2     &   2    &\\
2&  1.0    &  $0.125\times10^{5}$     &   0.5   &   0.2     &   2    &\\
3&  1.0    &  $1.0\times10^{5}$     &   0.125 &   0.03    &   2    &\\
4&  1.0     &  $1.0\times10^{5}$     &   0.125 &   0.05    &   2    &\\
5&  2.0     &  $1.0\times10^{5}$     &   0.125 &   0.05    &   2    &\\
6&  3.0     &  $1.0\times10^{5}$     &   0.125 &   0.05    &   2    &\\
7&  4.0     &  $1.0\times10^{5}$     &   0.125 &   0.05    &   2    &\\
8&  5.0     &  $1.0\times10^{5}$     &   0.125 &   0.05    &   2    &\\
9&  5.06  &  $1.0\times10^{5}$     &   0.125 &   0.05    &   2    &\\
10&  5.25    &  $1.0\times10^{5}$     &   0.125 &   0.05    &   2    &\\
11&  5.5    &  $1.0\times10^{5}$     &   0.125 &   0.05    &   2    &\\
12&  5.75   &  $1.0\times10^{5}$     &   0.125 &   0.05    &   2    &\\
13&  6.0    &  $1.0\times10^{5}$     &   0.125 &   0.05    &   2    &\\
14&  6.5    &  $1.0\times10^{5}$     &   0.125 &   0.05    &   2    &\\
15&  0.0     &  $0.625\times10^{4}$     &   2 &   0.8    &   2    &\\
16&  1.0     &  $0.625\times10^{4}$     &   2 &   0.8    &   2    &\\
17&  1.5     &  $0.625\times10^{4}$     &   2 &   0.8    &   2    &\\
18&  2.0     &  $0.625\times10^{4}$     &   2 &   0.8    &   2    &\\
\hline
\end{tabular}
\label{tbl:2}
\end{table*}

\section{Results}
In this section, we present the results of the numerical experiments.
In order to understand the evolutionary process more comprehensively,
we carried out a set of numerical experiments. The parameters and their
values for the experiments are listed in Table \ref{tbl:2}. Totally 18
cases are investigated, in which two correspond to the stable equilibrium
(cases 1 and 9), and the rest correspond to the nonequilibrium.

\begin{figure}
\centering
\includegraphics[width=10cm,clip,angle=0]{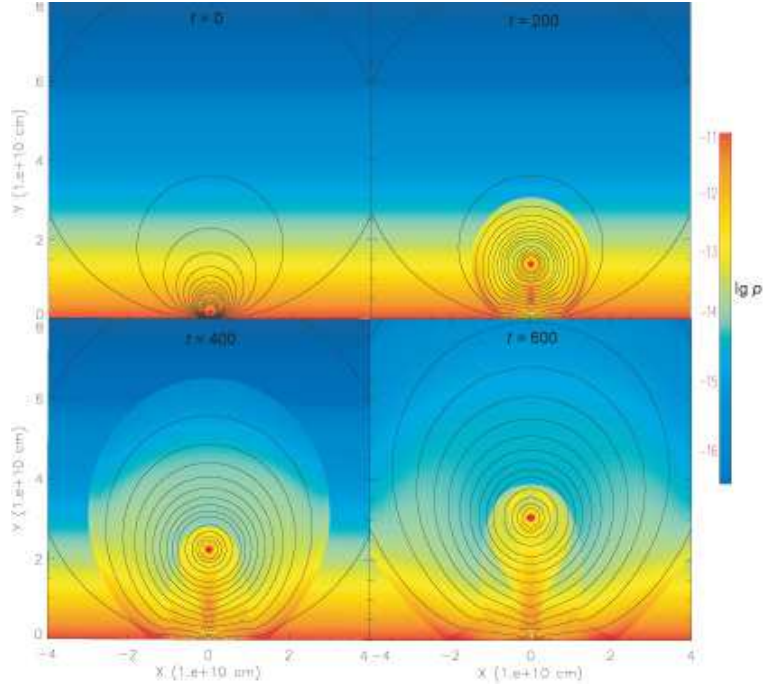}
\caption{Evolution of the magnetic field (black contours) and the plasma density (shadings) as
the eruption progresses for case 16. Propagation of the fast-mode shock around the flux rope is clearly seen.
And the X-type neutral point is distinct at \emph{t} = 600 s. The unit is second. The right color bar
represents values of the density in $\lg\rho$~(g~cm$^{-3})$.}
\label{fig:ms1880fig1}
\end{figure}

Now we take case 16 as an example to present evolution progresses of nonequilibrium from its initial state.
Figure \ref{fig:ms1880fig1} illustrates the evolution of the magnetic field
and the plasma density as the eruption progresses for case 16. Black solid lines represent
the magnetic field lines and shadings show the density distribution.
In this case, the initial state is not in equilibrium. Because the
magnetic compression outstrips the magnetic tension, the flux rope begins to rise quickly
from the start of the experiment. The closed magnetic field lines become stretched
with the lift-off of the flux rope, and the X-type neutral point appears on the boundary
surface in the magnetic configuration with time going on. This magnetic topology means the
magnetic reconnection occurs, i.e. there exits magnetic diffusion, which can convert magnetic
energy to heating and the kinetic energy of the plasma. In our experiments, although no
physical diffusion is included in equations (1), the results of numerical diffusion is equivalent
to the result of the physical diffusion (e.g. see detailed discussions given by Wang et al.
2009). Moreover, we can notice in these panels that propagation of the fast shock,
which is a crescent feature around the flux rope (Wang et al. 2009 and Mei et al. 2012a).

\subsection{Relation of equilibrium height of flux rope to the relative strength of background field}

In this section, we study the final height of the flux rope in equilibrium
as a function of the relative background field $M$ (i.e. the ratio of the
strength of dipole field $m$ and the product of the filament current $I$
and the depth $d$ of the dipole field). Since this study is based on the
equilibria curve through theoretical analysis as in Forbes (1990), we choose
the same radius and initial height of the flux rope, except the different
values of the parameter $M$ for cases 4-12. According to Equation (3) of
Forbes (1990), the flux rope is in the stable equilibrium  for only case 9.
Cases 3-8 and 10-12 are in the nonequilibrium at the initial time. On the
basis of Equation (3) of Forbes (1990), $M = 5.06$ is a critical point,
i.e. when $M < 5.06$, the final flux rope height is higher than the initial
height; when $M > 5.06$, the final height is lower than the initial one.
However, in our numerical experiment, the value of the critical point
becomes $M \approx 5.25$.

\begin{figure}
\centering
\includegraphics[width=15cm,clip,angle=0]{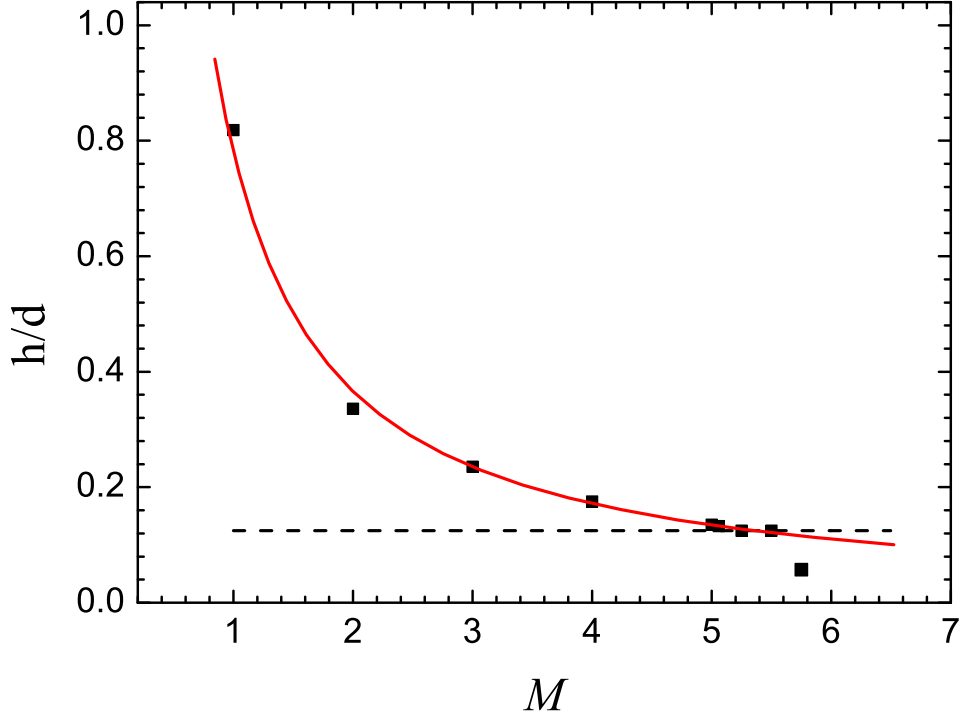}
\caption{The final height of the flux ropes as a function of the relative dipole magnetic field strength $M$.
Solid points indicate the final location of the flux rope. The dashed line denotes the initial hight of the
flux rope at $h/d = 0.125$. The red solid curve is a freehand interpolation of the numerical results, and it
displays the power-law function $h/d = 10^{-0.1}M^{-1.1}$, where $d$ is the depth of the background dipole field,
and $M$ is a dimensionless parameter which gives the relative strength between the dipole and the filament current.}
\label{fig:ms1880fig2}
\end{figure}

In order to get visual information, we plot Figure \ref{fig:ms1880fig2}, which
displays the final height of the flux rope as a function of the relative
dipole strength $M$. Solid points indicate the final flux rope location.
The starting point of the flux rope is at the location $h/d = 0.125$.
The dashed line is for the initial height. The red solid line is a
 fitting curve of the numerical results, which shows the power-law function
 $h/d = 10^{-0.1}M^{-1.1}$. From Figure \ref{fig:ms1880fig2}, we can see that
 the final height of the flux rope is a power-law function of the relative
 strength of the dipole field $M$ when $M \leq 5.25$. As $M \leq 5.25 - 0.3$,
 the location at which the flux rope stops is higher than the starting location
 of the flux rope. However, when $M \geq 5.25 + 0.3$, the final flux rope location
 is lower than the starting location of the flux rope. For $M = 5.75$ the stopping
 height of the flux rope is about 0.057, less than the initial height 0.125.
 So at $M = 5.25 \pm 0.3$, there appears to be a transition in the height at
 which the flux rope stops. The transition from upward to downward motion takes place
 at about $M\approx5.25$, which is close to $M = 5.06$ predicted by the vacuum
 equilibrium model for a filament or prominence of radius $0.05d$ (see Equation (3)
 of Forbes 1990).

Our results indicate that, from $M\approx5.0$ to $M\approx5.25$, the final
location of the flux rope is gradually changing, rather than steeply changing
in Forbes (1990). This is probably because of the much higher resolution in
our simulations, resulted from the double grid and the increase of grid points
in ZEUS-2D MHD code.

\subsection{Evolution of the system with different background fields}

In this section, we focus on the influence of the background field
on the evolution of the magnetic system. First, in order to solve
a few open questions in the work of Forbes (1990) and compare our
results with his. We investigate the influence of the different $M$
on the evolution of the system. The cases 15-18 are presented with
different $M$ given the same $d$, $h_{0}$ and $r_{0}$.

\begin{figure}
\centering
\includegraphics[width=15cm,clip,angle=0]{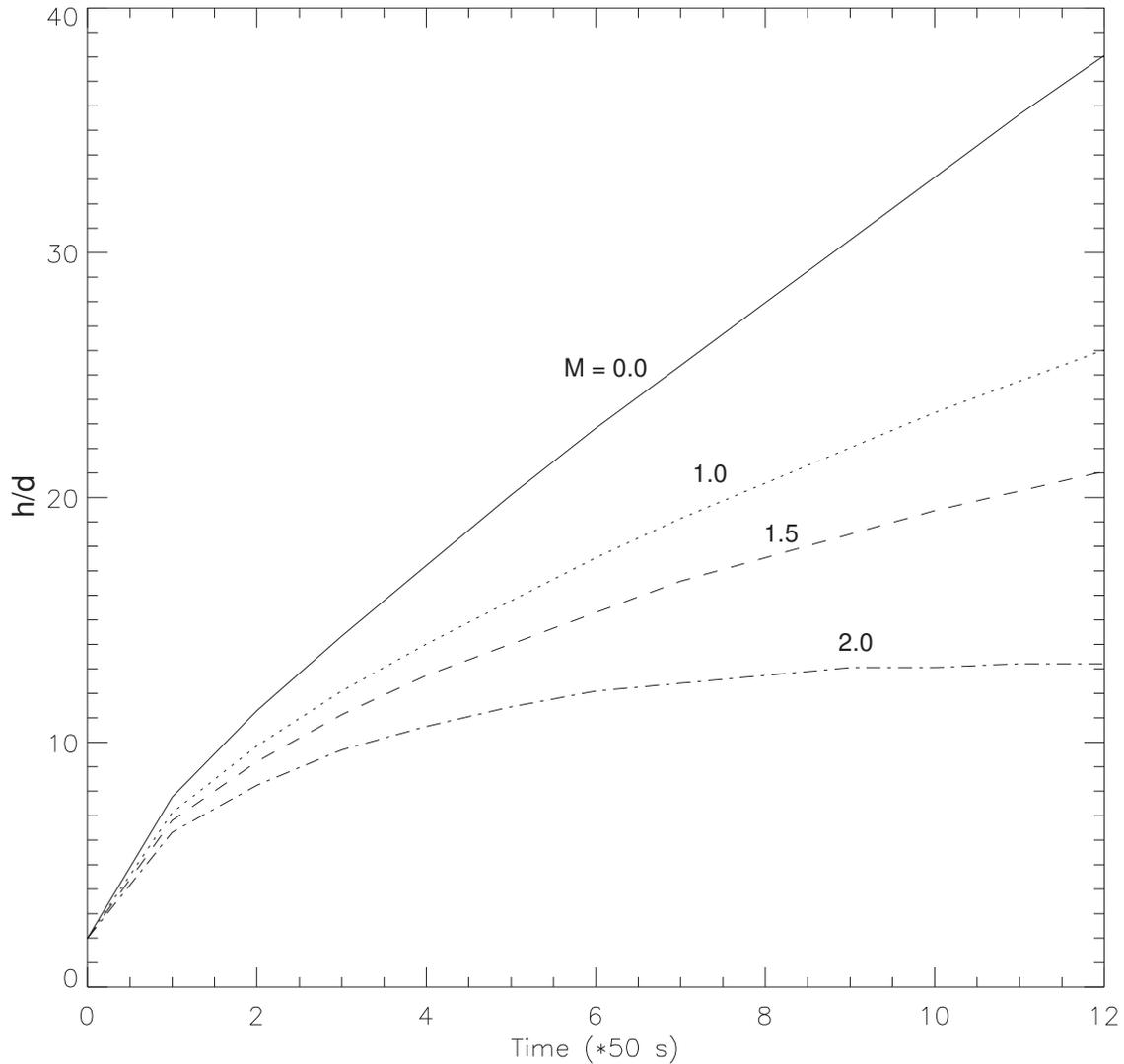}
\caption{$h/d$ as a function of time. $h$ is the flux rope height. $d$ is the depth of the background dipole field.
$M$ is a dimensionless parameter which gives the relative strength between the dipole and the filament current.}
\label{fig:ms1880fig3}
\end{figure}

Figure \ref{fig:ms1880fig3} plots the height of the flux rope $h$ as a function
of time in different $M$. When the background field is equal to zero ($M=0.0$),
the initial repulsive force on the flux rope is large, and the flux rope
promptly rises at the beginning. This agrees with the result in Forbes (1990).
However, after about $t=2.0$ s, our flux rope's trajectory differs from the
result of Forbes (1990). Figure 8 of Forbes (1990) shows that at about $t=2.0$ s
and $M=0.0$ the speed of flux rope becomes warped. However, this warp do not
represent in our simulations. The reasons why we have different results are:
1) this increase could be a numerical artifact of the open boundary conditions;
or 2) it is because of the lack of a gravitationally stratified solar atmosphere.
In this present work, we used the same boundary conditions of Forbes (1990) and
considered the gravitationally stratified background solar atmosphere. By comparison
of our numerical simulations with Forbes (1990), we find that the warp of the
height of flux rope may be eliminated by gravitational stratification effect.
The flux rope when $M = 1$ in the work of Forbes (1990) stops at some
height after it rises up at the very start, and then moves downward
slightly before continuing to rise further. Our results indicate that
the gravitationally stratified medium can account for this phenomenon.

At the beginning of the experiment, the flux rope keep rising
rapidly until the magnetic tension produced by the stretching
of the line-tied field lines becomes large enough to slow down
its upward motion. The Lorentz force $\mathbf{J}\times\mathbf{B}$
plays a main role in the decrease of the initially upward
velocity of the flux rope.
From Figure \ref{fig:ms1880fig3}, we see that for $M = 2$ the height
of the flux rope remains almost constant after $t = 500s$. It
is because that the flux rope reaches the new equilibrium state
at that time. By Equation (3) of Forbes 1990, we can further check
whether the state of the flux rope is in equilibrium or not. We
take the height of the flux rope $h$, the depth of the dipole $d$
and the relative strength of the dipole $M$ at $t = 600$ s into
Equation (3) of Forbes 1990, and find that the left almost equals
the right of Equation (3). This means that the new equilibrium state
is achieved.

\begin{figure}
\centering
\includegraphics[width=10cm,clip,angle=0]{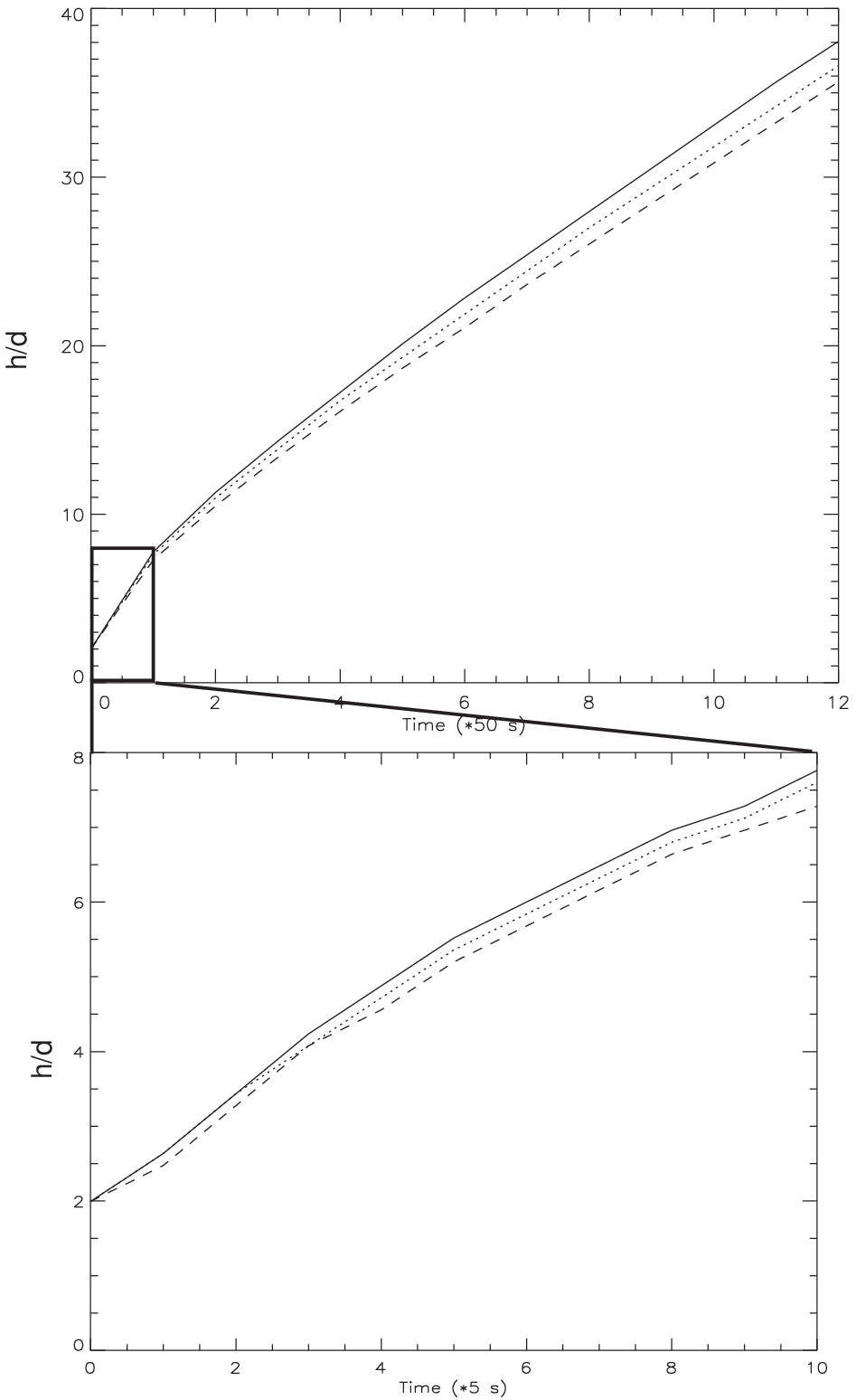}
\caption{$h/d$ as a function of time for the different background field $m$. $h$ is the height of the flux rope.
$d=0.625\times10^{4}$~km and $I_{0}=3\times10^{11}$~A are the the depth of the background dipole field and
the initial current strength of the flux rope. The solid line
corresponds to $m=0$, the dot curve is for $m=dI_{0}$, and the dashed curve is for $m=2dI_{0}$.
The lower panel is the zoom-out for time from 0 to 50 s.}
\label{fig:ms1880fig4}
\end{figure}

Since the evolution of the system in the corona may be
controlled by the background field, we need to investigate
how the evolution of the magnetic system relies on the
values of the background field strength, i.e. $m$.
Figure \ref{fig:ms1880fig4} shows the height of the flux rope
in the different background field strength $m$. The depth
of the background dipole field and the initial current
strength of the flux rope are $d=0.625\times10^{4}$ km and
$I_{0}=3\times10^{11}$ A. The solid line corresponds to $m=0$,
the dot curve is for $m=dI_{0}$ and the dashed curve for
$m=2dI_{0}$. From Figure \ref{fig:ms1880fig4}, we can see that
the height of the flux rope becomes higher when the background
field strength gets smaller. This implies that the flux
rope can escape more easily following the catastrophe if
the background field is weak.

\begin{figure}
\centering
\includegraphics[width=10cm,clip,angle=0]{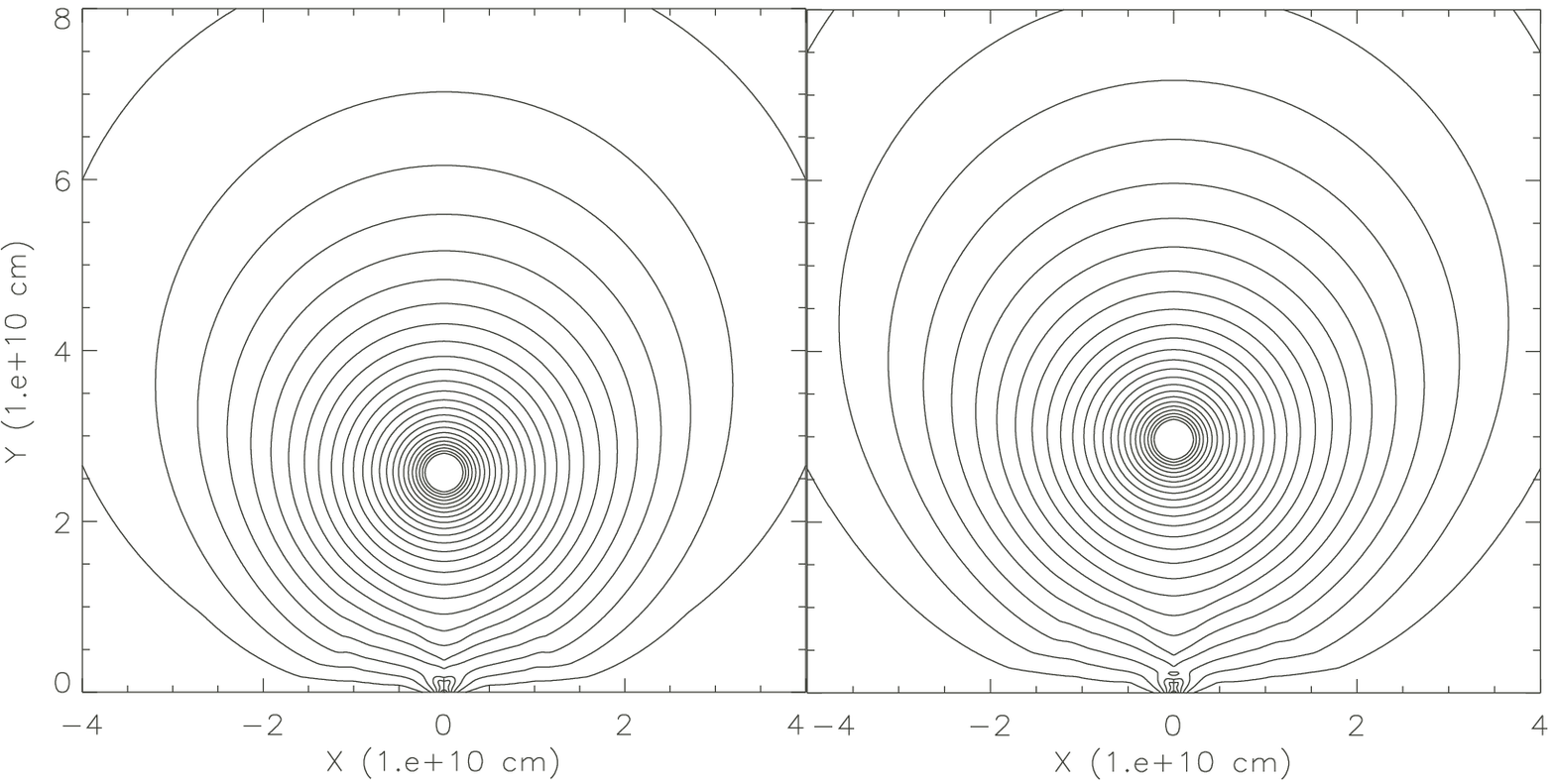}
\caption{Evolutions of the magnetic configuration at the different value of $m$ at $t=600$~s.
The left and right panels show $m=2dI_{0}$ and $m=dI_{0}$. They correspond to the dashed and dot
curves in Figure \ref{fig:ms1880fig4}, respectively.}
\label{fig:ms1880fig5}
\end{figure}

To further demonstrate the relation between the strength
of the background field and the flux rope, we studied the
evolutions of the magnetic configuration in two cases in
Figure \ref{fig:ms1880fig4}. As shown in Figure \ref{fig:ms1880fig5},
the magnetic field lines are represented by continuous
contours. The left and right panels show $m=2dI_{0}$ and
$m=dI_{0}$, and they correspond to the dashed, dot curves
in Figure \ref{fig:ms1880fig4}, respectively. From Figure
\ref{fig:ms1880fig5}, we can notice that the flux rope in
the right panel is higher than in the left one. In addition,
we are also able to recognize the existence of the X-point
in these two panels, which may result in fast energy dissipation
via magnetic reconnection.

\subsection{The internal evolution of the flux rope and effect of its radius}

\begin{figure}
\centering
\includegraphics[width=10cm,clip,angle=0]{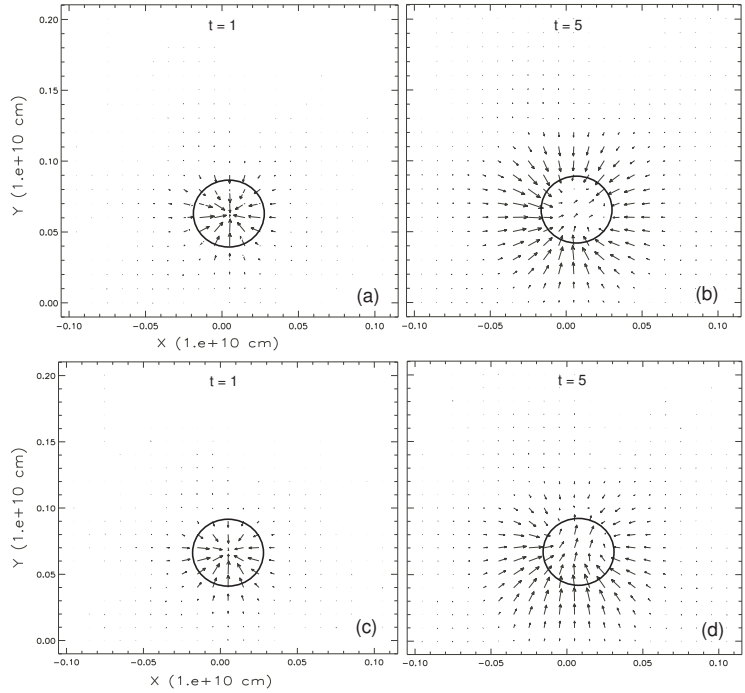}
\caption{Velocity streamlines at two different times for the stable equilibrium with $M=2.25$ (case 1) in (a) and (b),
and the nonequilibrium case with $M=1.0$ (case 2) in (c) and (d). The circles indicate the position of the flux rope.}
\label{fig:ms1880fig6}
\end{figure}

The flux rope moves upward very quickly driven by the
unbalanced magnetic compression at the beginning
of our simulation, whilst small perturbation on the
amplitude of the flux rope along its radial direction
always occurs since the initial state within the filament
is never in exact equilibrium. Flow always appears
within the filament almost at once as shown in Figure \ref{fig:ms1880fig6}.
The upper panel in this figure shows velocity streamlines
at two different times for the stable equilibrium with $M=2.25$
(case 1), while the lower panel shows velocity streamlines for
the nonequilibrium case with $M=1.0$ (case 2). The circle in each
panel represents the position of the flux rope. At $t=1$~s, the
flow speed for the stable case is about 0.8 the speed for the
nonequilibrium case. Meanwhile, the flow speed at $t=2$~s for the
stable equilibrium equals approximately the speed at $t=5$~s for
the nonequilibrium. These results show that the internal flow
of the flux rope remains small and does not last very long when
the initial state of the flux rope commences from the stable
branch of the theoretical equilibrium curve.

\begin{figure}
\centering
\includegraphics[width=10cm,clip,angle=0]{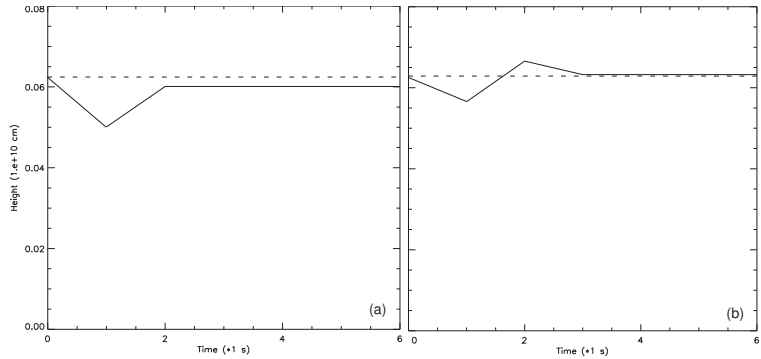}
\caption{The flux rope height $h$ as a function of time for the same stable equilibrium (case 1)
in the different computational domain with the same grid points: the left panel represents $(-4L, 4L)\times (0, 8L)$
with $800\times 800$ grid points, and the right panel depicts $(-L, L)\times (0, 2L)$ with $800\times 800$ grid points.
The solid curves are for case 1, and the dashed lines correspond to the initial height $h_{0}=0.0625\times10^{5}$~km for case 1.}
\label{fig:ms1880fig7}
\end{figure}

In order to understand the influence of the computational
domain on the internal evolution of the flux rope, we also
investigate the internal evolution of the flux rope in two
different computational domains. Figure \ref{fig:ms1880fig7} shows
evolution of the height of the flux rope with respect to
time for the stable equilibrium with $M=2.25$ (case 1) in
the two computational domains [$(-4L, 4L)\times (0, 8L)$
in (a) and $(-L, L)\times (0, 2L)$ in (b)] with the same
grid points $800\times 800$. The solid curves are evolution
of the height of the flux rope with respect to time for
case 1, and the dashed lines correspond to the initial height
$h_{0}=0.0625\times10^{5}$ km for case 1. We can see that the
readjustment of the height of the flux rope accomplished by
$t=2$ s in (a). After 2 s, the flux rope remains stationary
at the height about $0.06\times10^{5}$~km. However, after
taking $t=3$~s in (b), it becomes stationary at the height
about $\sim0.0625\times10^{5}$~km.

The information revealed by Figure \ref{fig:ms1880fig7} suggests that
the numerical diffusion is faster when the computational domain
is larger. Whereas, the numerical error is larger in (a) than
(b), since the numerical error can be estimated by the ratio
of the grid spacing to the initial height $\bigtriangleup x/h_{0}$,
i.e. 16\% in (a) and 4\% in (b). Because of the numerical error,
the stationary height of the flux rope is closer to the initial
height in (b) than in (a).

\begin{figure}
\centering
\includegraphics[width=10cm,clip,angle=0]{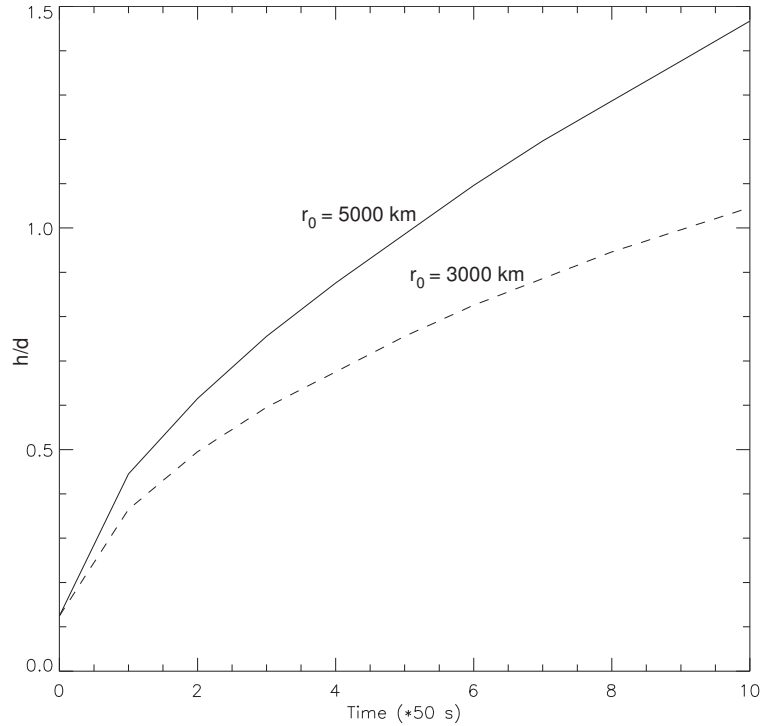}
\caption{Variation of the flux rope height versus time for cases 3 and 4: curves $r_{0} = 3000$ km is for case 3,
and curve $r_{0} = 5000$ km is for case 4. $d$ is the depth of the background dipole field and it is same for these two cases.}
\label{fig:ms1880fig8}
\end{figure}

In order to investigate the influence of the radius of flux
rope on its evolution, we have performed two simulations
(cases 3 and 4). We vary the radius of the flux rope in these
two cases, while other parameters remain unchanged.
The values of the parameters are listed in Table \ref{tbl:2}.
Figure \ref{fig:ms1880fig8} displays evolution of the height of
flux rope with respect to time in these cases. Curve
$r_{0} = 3000$ km is for case 3, while curve $r_{0} = 5000$
km is for case 4.
From this figure, we see that the flux rope with larger radius
apparently has faster upward velocity than that with smaller
radius, which means that greater radii can result in eruption
of flux rope more easily.

\section{Discussions and Conclusions}

We numerically investigate the evolution of the flux
rope using the ZEUS-2D code for modelling the prominence
or the filament in the corona, which may eventually erupt
as catastrophe. The empirical S\&G atmosphere model
is employed for the distribution of the density of the
background field.
Our present simulations has a higher resolution than the
previous work, e.g. Wang et al. (2009) and Forbes (1990),
due to the larger simulation domain and more grid points.
We studied the influence of the strength of the background
field and the radius of the flux ropes on the internal,
overall equilibrium and escape of the flux ropes in the
detailed simulations, including $18$ cases for the different
combinations of several important parameters. The main
conclusions are drawn as follows.

1. In our simulations, by using the realistic
plasma environment and much higher resolution,
we notice some different characters compared to previous studies in Forbes (1990).
We find that the speed of the flux rope
do not become warped after $t=2.0$ s for $M=0$ and for $M=1$
($M$ is the ratio of the strength of the dipole field $m$ and
the product of the filament current $I$ and the depth $d$ of
the dipole field, i.e. $M=m/(Id)$), which differs from the
results in Forbes (1990). The flux rope would rather keep
rising slowly, and stop at some height after some time, then
moves downwards slightly before continuing to rise further.

2. Among cases 4-12, the final height of the flux rope
varies with $M$ (the ratio of the strength of the dipole
field $m$ and the product of the filament current $I$ and
the depth $d$ of the dipole field, i.e. $M=m/(Id)$) in
the way of power-law function $h/d = 10^{-0.1}M^{-1.1}$.

3. The flux rope can escape more easily if the background magnetic
field is weaker. This implies that the catastrophe behavior can be
triggered by suppressing the strength of the background magnetic field,
which is consistent with previous work by Forbes (1990),
Isenberg et al. (1993), Lin et al. (1998, 2007), Chen (2011).
The decay of the photospheric magnetic field due to the magnetic
diffusion may result in the eruption of the flux rope (Mackay \& van Ballegooijen 2006),
and further explain why the peak rate of the CME occurrence is usually
delayed by $6-12$ months with respect to the peak of the Sunspot number
(Robbrecht et al. 2009).

4. The initial radius of the flux rope may have significant influence on its evolution.
The results indicate that the flux rope with larger initial radius erupts more easily.

5. The internal flow of the flux rope remains small and
does not last very long when the initial state of the flux
rope commences from the stable branch of the theoretical
equilibrium curve. We also find that the time and velocity
of this flow are related to the computational domain.
Provided that the grid points remain unchanged, the increase
of the computational domain can result in shorter time for
the internal equilibrium of the flux rope, whilst the numerical
error is in an expected range.

The authors appreciate C. Shen and Z. Mei for valuable discussions on the techniques of numerical simulation.
They are also grateful to the referee for valuable comments and suggestions that improved this paper.
This work was supported by the National Basic Research Program of China (2012CB825600 and 2011CB811406), and the Shan-dong Province Natural Science Foundation (ZR2012AQ016).
JL's work was supported by the Program 973 grants 2011CB811403 and 2013CBA01503, the NSFC grants 11273055 and 11333007,
and the CAS grants KJCX2-EW-T07 and XDB09040202.

\label{lastpage}

\end{document}